\def\be {\begin{equation}}
\def\ee {\end{equation}}
\def\p {\partial}

\documentstyle[preprint,eqsecnum,aps]{revtex}
\begin{document}
\draft
\preprint{UCSBTH 94-51, gr-qc/9412016}
\title{Almost Ideal Clocks in Quantum Cosmology:
A Brief Derivation of Time}
\author{Donald Marolf}
\address{Physics Department, The University of California,
Santa Barbara, California 93106} \date{November, 1994}
\maketitle

\begin{abstract}

A formalism for quantizing time reparametrization invariant dynamics
is considered and applied to systems which contain an `almost ideal clock.'
Previously, this formalism was successfully applied to the Bianchi
models and, while it contains no fundamental notion of
`time' or `evolution,' the approach does contain a notion of correlations.
Using correlations with the almost ideal clock to introduce a notion of
time, the work below derives the complete formalism of external time
quantum mechanics.  The limit of an ideal clock is found to be closely
associated with the Klein-Gordon inner product and the Newton-Wigner
formalism and, in addition, this limit is shown to fail for a clock that
measures metric-defined proper time near a singularity in Bianchi models.

\end{abstract}
\pacs{}

\section{Introduction}

The goal of this work is to {\it derive} the usual framework
of external time quantum mechanics from the what
we will call the `almost local formalism' of
\cite{QORD} for the quantization of time reparametrization invariant
finite dimensional systems.  By `external time quantum mechanics'
we mean the mathematical structure of the
Schr\"odinger picture consisting
of a Hilbert space ${\cal H}_{ext}$ of states labeled by a
parameter $\tau$ together with an algebra
generated by some set $L_i$
of Hermitian operators on ${\cal H}_{ext}$.  The operators should satisfy
\begin{equation}
\label{lie}
[L_i, L_j] = C^k_{ij} L_k
\ee
for some structure constants $C_{ij}^k$ and evolution
in the parameter $\tau$ is given by
\be
-i \hbar {\p \over {\p \tau}} |\psi(\tau)\rangle = H(\tau)
|\psi(\tau)\rangle
\ee
for some self-adjoint operator $H(\tau)$.  Equivalently, we
could derive a Heisenberg picture but we will see that construction
of the Schr\"odinger picture is most direct.
For a given system with a
classical analogue, the commutation relations \ref{lie} should
agree with the Poisson Bracket relations of the corresponding phase
space functions.

A derivation of this structure will
be possible despite the fact that the almost local formalism contains
no fundamental notion of time or evolution.  This approach
 is an implementation
of the ideas expressed by, for example DeWitt \cite{Bryce,tril}
and Rovelli \cite{Carlo,Carlo2,Carlo3,Carlo4}
that such features are unnecessary and,
in fact, inappropriate
for a description of time reparametrization invariant systems.  Thus,
we address the problem of time in quantum gravity.

The main ideas employed below are not new.  Actually, they are nearly
as old as (canonical) quantum gravity itself as we follow the basic
approach outlined in \cite{tril} and extended in {\it e.g.}
\cite{Banks,Jim,JH,CK1,CK2,PW}.
That is, we first use a method
based on Dirac's \cite{PAMD} to quantize the system by
imposing a Hamiltonian constraint on physical states and we
then use a semiclassical approximation to
study the correlations so induced between
some degree of freedom (which we choose to call a clock) and
the rest of the system.  We also follow
the suggestion of \cite{Bryce,Carlo,Carlo2,Carlo3,Carlo4,Lee} that
these correlations be captured through the use of
time reparametrization
invariant quantum operators.  The construction of these
operators will be based on the classical
methods of \cite{Bryce}.

The new input here
lies in bringing together the above ideas with the
technology of \cite{QORD} and its specific implementation of these
suggestions in the quantum context.  This provides
for a much more complete analysis than was previously possible
as we will construct the explicit
map between the physical Hilbert space
${\cal H}_{phys}$ of the almost local formalism and the usual
Hilbert space ${\cal H}_{ext}$ of external time quantum mechanics
while avoiding the difficulties \cite{KK} usually associated with such
derivations.
The complete formalism also forces us to take into account new subtleties
but leads in the end to a physically reasonable conclusion.

While a satisfactory formalism for full
quantum gravity remains out of reach, the
almost local approach was shown in \cite{QORD} to provide
a complete quantization of Hamiltonian vacuum Bianchi models
(see, e.g. \cite{bj,atu,RySh}) and of similar finite dimensional
time reparametrization invariant systems.  In particular, it
was successfully applied \cite{BIX}
to the complicated mixmaster cosmology \cite{MTW}.
The details of this approach will be reviewed in section \ref{prelim}
but for now we mention that it provides \cite{QORD,BIX} a
positive definite inner product on a physical Hilbert space ${\cal
H}_{phys}$ as well as a complete set of well-defined symmetric
quantum {\it observables} on this space (perennials in the language of
\cite{KK}) that, in a certain sense, capture
the notion of `the value of quantity $A$ at the point in time at
which quantity $Q$ takes a specified value $\tau$.'  In addition,
this approach captures in a quantum sense the classical recollapsing
behavior of the appropriate cosmological models.

After reviewing the almost local formalism in section \ref{prelim}, we
describe in section \ref{approx} the class of systems to be
considered and the approximations
to be used.  In particular, we introduce
the notion of an `almost ideal clock.' We then derive the
external time formalism in three stages in section \ref{derive}.
We begin by showing in \ref{unit} that `evolution' in the clock time $\tau$
is unitary to the extent that the clock is ideal and then show
in \ref{alg} that our observables have the usual algebraic structure
associated with external time quantum mechanics (in particular, that
they satisfy the canonical commutation relations).  Section
\ref{map} derives the explicit map between ${\cal H}_{phys}$
and ${\cal H}_{ext}$ and shows that it is of the form associated
with the Klein-Gordon inner product and the Newton-Wigner operators.
Section \ref{diss} points out certain interesting features of our
derivation such as the fact that, for clocks that measure
metric-defined proper time, our
approximations always fail near a singularity in homogeneous models.

\section{Preliminaries}
\label{prelim}

We now embark on a brief review of the almost local formalism of
\cite{QORD} for the reader's convenience.
This formalism is a refinement
of the original prescription\cite{PAMD} of Dirac for the canonical
quantization of constrained systems, combined with a suggestion
of DeWitt\cite{Bryce} for the construction of observables
(perennials in the language of \cite{KK}).
A more detailed description which takes a geometric approach
toward the classical phase space and an algebraic approach toward
the quantum operators was presented in \cite{QORD} and the appropriate
derivations may be found there.

For convenience, we
consider systems for which
time reparametrization is the only gauge symmetry.  In a classical
Hamiltonian formalism, such systems are described by a phase space
$\Gamma$ and a constraint $h = 0$ where $h$ is some function on
$\Gamma$.  Also for simplicity, consider the case
$\Gamma = T^*{\bf R}^n$ with coordinate functions $p_i,q^j$
for $i,j$ in some index set $I$.  As usual, we choose these to
satisfy the Poisson bracket
algebra $\{q^i,p_j\} = \delta_i^j$.  The quantization
procedure of \cite{QORD} can be summarized by introducing the usual
operators $P_i,Q^j$ on the Hilbert space ${\cal H}_{aux}
= L^2({\bf R}^n,d^nq)$ where
the $Q^j$ act by multiplication and the $P_i$ by derivation with
the commutation relations $[Q^i,P_j]=i\delta_i^j$.  Similarly, a self
adjoint operator $H$ on ${\cal H}_{aux}$ is to be constructed from some
appropriately factor ordered version of the constraint function
$h(p,q)$.  As in \cite{QORD,BIX}, we take capital letters to
refer to quantum operators and lower case letters to refer to their
eigenvalues and classical counterparts.

Next introduce a
one parameter family of `lapse operators' $N(t)$
that are proportional to the identity ${\openone}$ on ${\cal H}_{aux}$.
That is, $N(t) = n(t) \openone$ for some $n: {\bf R} \rightarrow
{\bf R}$.  We require $n(t)$ to be continuous and to satisfy
the boundary conditions
\begin{equation}
\int_0^{\pm \infty} n(t) dt = \pm \infty
\end{equation}
but physical operators will be independent of which $n(t)$ is
chosen.
For each classical function $q^j$ and $p_i$, we now introduce a
one parameter family of operators through
\begin{eqnarray}
Q^j(t) \equiv \exp[i \int_0^tdt N(t) H] Q^j \exp[-i \int_0^tdt N(t) H]
\cr
P_i(t) \equiv \exp[i \int_0^tdt N(t) H] P_i \exp[-i \int_0^tdt N(t) H].
\cr
\end{eqnarray}

Following \cite{PAMD}, we will construct a physical Hilbert space
by `imposing the constraint' $H|\psi_{phys}\rangle = 0$
on `physical states $|\psi_{phys}\rangle$.'  Precisely, we assume that
the Hilbert space ${\cal H}_{aux}$ has a basis $\{|n,E\rangle :n\in {\bf Z},
E \in [a,b]\}$ for the part of ${\cal H}_{aux}$ corresponding to some
spectral interval $[a,b]$ of $H$ with $0 \in [a,b]$.  These
states are to satisfy
\begin{equation}
\label{aux}
\langle n_1,E_1 | n_2,E_2 \rangle = \delta_{n_1,n_2}  \delta (E_1 -
E_2), \ H|n,E\rangle = E|n,E\rangle
\end{equation}
and we thus require $H$ to have continuous spectrum that contains
the interval $[a,b]$ while having no point spectrum in this
interval.  We note that this is the case \cite{QORD,BIX}
for the proper formulation of
interesting cosmological models.
The physical Hilbert space ${\cal H}_{phys}$ may be defined
by stating that the formal symbols $|n,0\rangle$ span a dense
subset of ${\cal H}_{phys}$ and have the `spectral analysis inner
products' \cite{MG}
\begin{equation}
\langle n_1,0| n_2,0 \rangle = \delta_{n_1,n_2}.
\end{equation}
Note that this can be described as an infinite renormalization of
the `bare' inner product \ref{aux}\footnote{This inner product has been
independently introduced several times in related contexts.  See
\cite {HT,AH} for the cases known to the author.}.

Finally, we need to construct physical operators on ${\cal H}_{phys}$.
In general, we expect an operator $A$ on ${\cal H}_{aux}$ to define such a
physical operator if $[H,A]=0$.  It was shown in \cite{QORD} that a
large class of such physical operators $\Omega$ can be constructed from
the `time-dependent'  operators $\omega(t) = \omega(Q^j(t),P_i(t))$
through\footnote{Note that the notation of \ref{Om} differs
slightly from that in \cite{QORD}.}
\begin{equation}
\label{Om}
\Omega = \int dt N(t) \omega(t).
\end{equation}
Convergence of this integral was discussed in \cite{QORD,BIX} and
will not be considered here.  We simply use the result of
\cite{QORD} that, when  \ref{Om} converges, it defines a
physical operator $\Omega_{phys}$ on ${\cal H}_{phys}$
through
\begin{equation}
\label{mes}
\Omega_{phys} |n,0\rangle \equiv 2\pi \sum_m |m,0\rangle
\langle m,0 |\omega(t)|n,0\rangle_{aux}.
\end{equation}
The matrix elements on the right hand side are in fact independent
of $t$ and the
subscript $aux$ is a reminder that this matrix element
is to be evaluated in the auxiliary Hilbert space ${\cal H}_{aux}$, using
the corresponding inner product.  It is readily shown that the map
$\Omega \rightarrow \Omega_{phys}$ is a homomorphism; that is, for
$[A,H] = 0 = [B,H]$, we have $(AB)_{phys} = A_{phys} B_{phys}$ and
$(A+B)_{phys} = A_{phys} + B_{phys}$.  Thus, no harm is done by
dropping the subscript $phys$ from the operators on ${\cal H}_{phys}$.
We do so in order to simplify our notation; whether $\Omega$
represents the operator on ${\cal H}_{aux}$ or ${\cal H}_{phys}$ should be
clear from the context.

The particular operators
considered in \cite{QORD} were of the form \ref{Om} with
$\omega(t) =  \{A(t), i\hbar[\theta(B(t) - \tau),H]\}$.
The associated physical operators describe a certain sort of average
over `the
values of the quantity $A(t)$ at the times for which
$B(t) = \tau$,' even when the corresponding classical
quantity $b(t)$ may take on the given
value $\tau$ more than once along a trajectory.  However, when $b(t)$
takes this value many times the convergence of \ref{Om}
becomes a subtle issue so that our definition of an almost
ideal clock below must pay due attention to the corresponding classical
solutions.  Finally, note that the choice of $\Gamma = {\bf R}^{2n}$
and the use of canonical coordinates and momenta is not necessary, but
only a simple and familiar case.  As in \cite{Abook}, we may start with
any Lie algebra of overcomplete functions on $\Gamma$ and use any
unitary Hilbert space representation as the auxiliary space
${\cal H}_{aux}$.

\section{The Approximation Scheme}
\label{approx}

We now describe the systems to be considered and introduce the
approximations relevant to our derivation of external time
quantum mechanics.  That is, we describe what it means for a system
to contain an almost ideal clock and in what sense the ideal limit
is to be taken.  We also provide a number of useful tools for later
calculations.

\subsection{Almost Ideal Clocks}
\label{AIC}

The philosophy of this work is that we should, in principle, be
able to choose {\it any} degree of freedom to be a clock and thereby
{\it define} a notion of time.  Thus, we would like our definition
of an almost ideal clock to be as broad as possible.  Nevertheless,
we must recognize that some clocks are inherently better than others
in the sense that the notion of time they define is most like the
Newtonian one.  This is true even classically and so should come as
no surprise in the quantum context.  We thus restrict
our attention to a clock that, in some sense, interacts weakly
both with itself and with the rest of our system.

In order to define almost ideal clocks, it is useful to
first establish the concept of an {\it ideal} clock.  We
consider an ideal clock to be a system with a pointer whose
position $Q_p$ increments in direct proportion to the passage of
proper time.  Furthermore, the corresponding constant of proportionality
should be independent of the state of the system.  This definition is,
however, ambiguous for the systems we consider as there will, in
general, be two distinct notions of proper time.
One such notion is the proper time $\int_{t_1}^{t_2} N(t) dt$
associated with any system described by a
Hamiltonian constraint
$H$ and lapse $N$ while the other is whatever concept of proper
time may be defined by a metric for the system.  The point here
is that $N$ may not correspond exactly to the metric-defined lapse but
may be rescaled in some nontrivial way.  This is exactly
the case of interest as the almost local formalism can only be applied
(see \cite{QORD,BIX}) to homogeneous cosmological models
by using a constraint $H$ and lapse $N$ which are related to the
ADM constraint and lapse by $H = \sqrt{2 \over {3\pi}}
(\det g)^{1/2} H_{ADM}$ and $N = \sqrt{3 \pi}N_{ADM}/ \sqrt{2}
(\det g)^{1/2}$.  Here, $\det g$ is the determinant of the 3-metric on a
homogeneous spatial slice.

For the purposes of this paper we define an ideal clock to measure
the proper time associated with our choice of lapse and Hamiltonian
constraint and not the metric-defined proper time.  The rationale
for this choice is simply that it is in {\it this} case that we
may derive the usual external time quantum mechanics.  While this
state of affairs may seem unsatisfactory at first, section \ref{diss}
will $i)$ show that the notions of `almost ideal clocks'
associated with the two kinds of proper time
in fact agree unless the volume of the spatial slices is very small
and $ii$) point out that a clock which explicitly measures metric-based
proper time for arbitrarily small spatial volumes cannot even classically
define a satisfactory notion of time.

Thus, a time reparametrization invariant system with an ideal clock
is described by a constraint of the form
\be
\label{ideal}
0 = H = P_p + H_1
\ee
where $P_p$ is the momentum conjugate to $Q_p$ and $H_1$
is independent of $P_p$ and $Q_p$.
For such a system, \cite{QORD} has already shown that the external time
formalism can be derived {\it exactly}.  Our task is to show that when
the dynamics and quantum state conspire to create a regime in which the
clock is {\it almost} ideal, then evolution in the clock time is
described by the usual external time formalism to the same degree.

In general,
we expect a generic degree of freedom to approximate an ideal clock
when it is weakly interacting, both with other degrees of freedom and
with itself.  We therefore say that a reparametrization
invariant system contains a {\it decoupled} clock when the
Hamiltonian constraint may be written in the form
\be
\label{form}
0 = H = H_0 + H_1
\ee
where $H_0$ depends only on the coordinates and
momentum of the clock and $H_1$ is independent
of the canonical
clock variables.  While this situation is unphysical, we assume that
it may in turn be approximated by systems which differ
from \ref{form} only by terms that may be treated
adiabatically.  However, we will not consider this approximation here
and we take \ref{form} as our starting point.

We would like to treat the
case where $H_0$ is of the form $H_0 = H_p  = {{P^2} \over 2}
+ V(Q)$.  For such systems, we must
work in the approximation that the clock coordinate
$q$ is far from the classical
turning points so that the classical $q(t)$ is a single-valued
function of time.  Thus, we will be able to describe our clock
using a WKB approximation.

Now, if $V(q)$ confines $q$ to some finite region, what $H_p$
describes in not a clock but a {\it pendulum}; that is, a part of
a clock, or a clock that measures time only modulo
some (possibly state dependent) period.  Thus, specifying
the coordinate $Q$ determines not a unique moment of time but
a discrete set of such moments.  Even classically, such
a pendulum can only be used as a clock during some agreed upon
{\it half} of an oscillation (such as the right-moving half of the first
oscillation after the explosion of a firecracker).

Following the model of a grandfather clock, we will assume that this
ambiguity is removed by coupling the pendulum to some counter which
increments with each complete oscillation.  In contrast to the
grandfather clock, however, the primary part of our system will be the
{\it pendulum} and not the counter.  The counter will be used only to
specify the oscillation in which we are interested and, in what
follows, we use the right-moving half of the cycle in which the counter
reads zero.  Alternatively, we could couple the counter so that it
increments with each {\it half} of the oscillation, but the above choice
has the advantage that it
simultaneously considers the case where $V(q)$
has only one turning point (or none at all)
and {\it no} counter is present in the system (the counter is irrelevant
in this situation).  This is the
case when our clock is built from the scale factor of a homogeneous
cosmological model.  Because the counter will only be used in a
secondary manner, we expect the details of its construction and
coupling to the pendulum to be irrelevant so long as this coupling is
weak.

Nevertheless, for a consistent treatment it is useful to have an
explicit model of the counter.  We therefore take our clock
to be
described by the canonical pairs $(Q,P)$ for the pendulum and $(Q_c,P_c)$
for the counter with the Hamiltonian:
\be
\label{cc}
H_0 = {{P^2} \over 2} + V(Q) + \epsilon { {e^{-(Q-q_0(H_p))^2/4\lambda}
|P| P_c  e^{-(Q-q_0(H_p))^2/4\lambda}} \over \sqrt{4 \pi \lambda}}
\ee
where we have assumed that $V(q)$ has no more than a
single critical point and that this point (if it exists) is
a global minimum.  Here, $H_p = P^2/2 + V(Q)$, $\epsilon \ge 0$, and
$q_0(E)$ is the position of the left
classical turning point of a pendulum with energy $E$ when not coupled to the
counter; that is, the leftmost solution of $E = V(q)$.  Note that
\ref{cc} is constructed so that the counter coordinate $q_c$ advances
a distance $\epsilon$ with each oscillation of the pendulum.
Thus, we will say that our system contains an `almost ideal clock'
if it is described by a constraint which differs from \ref{form}
only by adiabatic factors and if $H_0$ describes a pendulum coupled
(when the pendulum is bound) weakly to a counter.

We now state precisely what is meant by
the ideal clock limit.  First, we assume that
$|\epsilon p_c|$ is small in the sense that we consider only states
$|\psi \rangle \in {\cal H}_{phys}$ whose spectral support in $P_c$
is concentrated in the region with $\epsilon p_c \ll \Delta E$, where
$\Delta E$ is the splitting between the energy levels of $H_p$ on which
the physical state $|\psi\rangle \in {\cal H}_{aux}$ is
concentrated\footnote{Here, we abuse notation as $|\psi\rangle$ is to
satisfy $H|\psi\rangle = 0$ and so is not a {\it normalizable}
state in ${\cal H}_{aux}$, but only in ${\cal H}_{phys}$.}.  Thus,
the pendulum-counter coupling is weak and the spectral projections of
$H_0$ and $H_p$ on $|\psi\rangle$ agree.
Choosing some pendulum position $\tau$, we
assume that for each eigenvalue $E$ of $H_0$ either $|\tau - q_t| \gg
\hbar / \sqrt{E}$ for any solution of $V(q_t) = E$ or that the
corresponding projection of $|\psi\rangle$ is negligible.
Thus, the pendulum is far form its turning points.
We also assume that the width $\lambda$ in \ref{cc} is large ($\lambda
\gg \hbar/ \sqrt{E}$)
so that the notion of a turning point need not be
defined too precisely.  Finally,
we ask that the projection of $|\psi\rangle$ to the subspace where
$Q = \tau$ have negligible projections onto the eigenvalues
$0$ and $\epsilon$ of $Q_c$ so that the counter may be
read clearly.

The observables associated with an almost ideal clock are
more complicated than those of an ideal clock as they depend on
the counter reading and the direction of
pendulum motion as well as the pendulum position.
In analogy with \cite{QORD,BIX}, we will capture
the notion of `the value of $A$ when the pendulum is moving to the
right through position $\tau$ and the counter reads $n$' with
the observables:
\be
\label{obs}
A^{\pm}_{\tau,n}  = {1 \over 2} \int_{-\infty}^{+\infty} dt \
\theta(\pm P(t)) \{A(t), \chi_n(t) {\p \over {\p t}} \theta(Q(t) - \tau)\}
\theta(\pm P(t)).
\ee
where $A(t) = A(Q^i(t),P_j(t))$ may be built from the coordinates
and momenta of both the clock and the other degrees of
freedom and $\chi_n(t)$ is the projection onto the spectral interval
$[n\epsilon, (n+1) \epsilon]$ of $Q_c(t)$.
We will not address the convergence of \ref{obs} here, although we
expect that, despite the more complicated form of \ref{obs} as compared
with the observables of \cite{QORD}, the same arguments may be applied
and that \ref{obs} converges to give a densely defined operator
on ${\cal H}_{phys}$ for a reasonable set of operators $A$.
Note that the commutator of $Q_c$ with ${\p \over {\p t}} Q(t)$
is of order $\epsilon p_c$ and so negligible.
Thus, these operators are
symmetric on ${\cal H}_{aux}$ to lowest order in $\epsilon p_c$.
Since we are only interested in the case where the counter takes the
value zero and
the pendulum moves to the right,  we drop
the labels $(n,\pm)$ on these operators with the understanding that
they always implicitly takes the values $(0,+)$.  Analogous results
follow when the
arguments below are applied to the remaining cases.

\subsection{The Physical States}
\label{states}

We now present the details of the approximation scheme to be
used in the rest of this work.  As mentioned above, by
assuming that the pendulum is far from its classical turning points, we
will be able to employ WKB techniques to great effect.  In
general, we will keep only terms of order $(\hbar)^0$ in our
approximation and discard higher terms.  However, we
will not discard {\it all} terms of order $\hbar$, but only those
associated with the WKB expansion.  Thus, we expect the discarded terms
to be small whenever the clock
is far from a turning point regardless of the behavior of the rest
of the system.  In fact, when the pendulum is confined between two
exponential walls, this approximation may be turned into an
expansion in $1/(\tau - q_t)$ (where $q_t$ is an associated classical
turning point) in the usual way.

As in \cite{QORD} (and in quantum mechanics generally),
we will find that it is essential to keep proper account of the
normalization of states in order to arrive at physically sensible
answers.  Because of the way that the inner product on ${\cal H}_{phys}$
is defined from that of ${\cal H}_{aux}$, we will
see that the presence of the counter is crucial to obtaining properly
normalized states.
It is this subtlety on which we
focus in our description of the physical states.

For the moment, we consider only the clock part of the the system and
make use of the factorization ${\cal H}_{aux} =  {\cal H}_0
\otimes {\cal H}_1$ where $H_0 = \hat{H}_0 \otimes \openone$
and $H_1 = \openone \otimes \hat{H}_1$ with respect to this
decomposition.  Since the physical `subspace' of ${\cal H}_{aux}$
is spanned by the states $|E_0,a_0,E_1,a_1\rangle = |E_0,a_0\rangle \otimes
|E_1,a_1\rangle$ with $\hat{H}_0|E_0,a_0\rangle = E_0|E_0,a_0\rangle$,
$\hat{H}_1|E_1,a_1\rangle = E_1|E_1,a_1\rangle$, and $E_0 = - E_1$,
($a_0$ and $a_1$ are degeneracy labels), we may concentrate
on the states $|E_0,a_0\rangle$.  The form of the
$|E_1,a_1\rangle$ will be unimportant.

In what follows, we will make use of the Hilbert space
${\cal H}_p \cong L^2({\bf R})$ associated with the pendulum and
the space ${\cal H}_c \cong L^2({\bf R})$ associated with the counter.
In general, the decomposition ${\cal H}_0 = {\cal H}_p \otimes_t
{\cal H}_c$ will depend on a choice of parameter time $t$.
Since \ref{mes} is independent of the choice of $t$, no harm is done by
choosing $t=0$.  We do so here and suppress the label $t$ on the
states and operators below.

We make no effort to label states explicitly with the
Hilbert space to which they belong nor, hereafter, do we distinguish
related operators that act on different spaces.  Thus, we drop the
hats on $\hat{H}_0$ and $\hat{H}_1$ and also refer to these operators
as $H_0$ and $H_1$.
In addition,we will make use of several bases for each Hilbert
space.
Both the space in which a state lives and the
basis to which it belongs are determined implicitly by the labels inside
the bra or ket.
For example, $|p_c\rangle$ denotes an eigenstate of $P_c$
in ${\cal H}_c$ with eigenvalue $p_c$ while $|q,p_c\rangle$
is an element of ${\cal H}_0$.  As mentioned above, the issue of
normalizations will be subtle and we will often make use
of pairs of bases whose members differ only in their normalizations.
In this case, one basis will be denoted in the usual way with
angle-bracket bra's $\langle |$ and ket's $|\rangle$ while
the other will be denoted by round bracket $(|$,$|)$ bra's
and kets.  Thus, $|p_c)$ and $|p_c\rangle$ are both
eigenstates of $P_c$ in ${\cal H}_c$ with eigenvalue $p_c$, but with
differing normalizations.  In general, the round-bracket basis will
be introduced first, but only the angle-bracket basis will be used in
the derivation of the external time formalism in section \ref{derive}.
The normalizations of bases will always be explicitly stated in the text.

Thus, we begin with a basis
$\{|\tau,p_c)\}$ of ${\cal H}_0$,
where $P_c |\tau,p_c) = p_c |\tau,p_c)$ and
$Q|\tau, p_c) = \tau |\tau, p_c)$ with
$(\tau, p_c | \tau',p'_c ) = \delta(\tau-\tau')
\delta(p_c - p'_c)$.
We note that $[P_c,H_0]
= 0$, so that, since the eigenstates of $H_0$ for a given value of $p_c$
are nondegenerate, we may label the $nth$ eigenstate of $H_0$
at $p_c$ by $|n,p_c) = |n;p_c) \otimes |p_c)$,
where $(p'_c|p_c) = \delta(p'_c - p_c)$ and
$|n;p_c)$ is the unit normalized
$n$-th eigenstate on ${\cal H}_p$
of the operator $H^{p_c}_0$ obtained from $H_0$ by replacing $P_c$ with the
eigenvalue $p_c$. The reader should beware of the difference
between $|n;p_c) \in {\cal H}_p$ and $|n,p_c) \in {\cal H}_0$ --
the former is the eigenstate of the operator $H^{p_c}_0$
(where $p_c$ determines just which operator this is!) while
the latter is the simultaneous eigenstate of $H_0$ and $P_c$.
Invoking the usual WKB approximation  \cite{WKB} and
assuming that $\epsilon p_c$ is small, such states satisfy
\begin{eqnarray}
\label{WKB1}
\langle \tau, p'_c  | n,p_c) &=& \delta(p'_c - p_c)
{{C(n,p_c)} \over {(2 \pi [2E(n,p_c) - 2V(\tau)]^{1/2})^{1/2}}} \Big[\alpha_1
\exp\biggl(i  \int_0^{\tau} \sqrt{2E(n,p_c) - 2V(q)} \ {{dq} \over \hbar}
\bigg)
\cr &+& \alpha_2\exp\biggl(-i  \int_0^{\tau} \sqrt{2E(n,p_c) - 2V(q)} \
{{dq} \over \hbar}
\bigg) \Big] [ 1+ {\cal O}(\epsilon) + {\cal O}(\hbar)]
\end{eqnarray}
where $C(n,p_c)$ is real and $\alpha_1,\alpha_2 \in {\bf C}$ have unit norm.
We expect this approximation to be valid when $V(\tau) \ll E(n,p_c)
\equiv (n;p_c|H_0^{p_c}|n;p_c)$.

In order to make use of \ref{mes}, we
must choose our basis $|E_0,a_0,E_1,a_1\rangle$ to satisfy
\ref{aux}.   In particular, it will be simplest to proceed by taking
these energy states to be delta-function normalized in the
total {\it clock} energy.  Note that the spectrum of $H_0$ is, in fact,
entirely continuous since we have
\be
\label{spread}
{{\partial E(n,p_c)} \over {\p p_c}} = (n;p_c| \epsilon
Z |n;p_c) > 0
\ee
where $Z$ is the coefficient of $\epsilon P_c$ in \ref{cc} and the inequality
follows because $Z$ is a strictly positive operator.  Thus, the
coupling to the counter broadens the pendulum energy levels into
continuous bands.  Note that our requirement that $|\epsilon p_c| \ll
\Delta E$ is just the assumption that these bands remain well
separated.

Now, the states $|n,p_c)$ are normalized according to
$(n,p_c|n',p'_c ) = \delta(p_c - p'_c)
\delta_{n,n'}$.
However, we would like to use the states $|n,p_c\rangle
=  ({{\p E(n,p_c)} \over {\p p_c}})^{-1/2} |n,p_c)$
which satisfy
\be
\label{norm}
\langle n, p_c| n' p'_c\rangle = \delta_{p_c,p'_c}
\delta(E(n,p_c) - E(n',p'_c)).
\ee
To derive the WKB approximation for $(\tau, p'_c|n,p_c\rangle$,
we evaluate the above derivative semiclassically using
\ref{spread}.  We begin by replacing the operator $|P|$ with the
value $\sqrt{2E(n,p_c) - 2V(\tau)}$.  Next, note that
$Z$ is peaked about a classical turning point in a Gaussian manner with
a width $\lambda$.  Since $\lambda \gg 1 / \sqrt{E(n,p_c)}$,
we may use our WKB approximation to evaluate
the contribution to \ref{spread} from the region
between the turning points and we may also
approximate $|\alpha_1 \exp(i f(\tau)) +
\alpha_2 \exp(-i f(\tau))|^2$ by its average value (which is $2$).
Here, $f(\tau) = \int_0^{\tau}
{{dq} \over \hbar} \sqrt{2E(n,p_c) - 2V(q)}$.

Because the exact wavefunction will decay exponentially
beyond $q_0(E)$, we multiply \ref{WKB1} by
$\theta(q - q_0(E))$\footnote{This assumes that $V$ increases to the left
at $q_0(E)$.  If $V$ has two turning points at $E$, this follows from
our definitions.  If $V$ has only one turning point at $E$, we
may use $\theta(q_0(E)-q)$ if $V$ increases to the right of $q_0(E)$.}.
Since $q_0(E)$ is the center of the Gaussian support of $Z$, the integral
over $q$ contributes a total value of $1/2$, canceling the
$2$ from the oscillatory terms.   It thus follows that
\be
\label{slope}
{{\p E(n,p_c)} \over {\p p_c}}  =  {{\epsilon C^2(n,p_c)} \over {2 \pi}}
\ee
to lowest order in $\epsilon$ and $\hbar$.  Distributing \ref{slope}
among the pendulum and counter factors and moving an $\hbar$
between them, the states of
interest satisfy $|n,p_c\rangle = |n;p_c\rangle \otimes
|p_c\rangle$ where
\be
\label{cn}
\langle q_c| p_c\rangle = {{e^{iq_cp_c/\hbar}} \over
\sqrt{\epsilon } }
\ee
and
\begin{eqnarray}
\label{WKB}
\langle \tau|n;p_c\rangle &=& {{\alpha_1} \over \sqrt{2 \pi \hbar
(2E(n,p_c) - 2V(\tau))^{1/2}}} e^{i\int_0^{\tau}  \sqrt{2E(n,p_c)
- 2V(\tau)}/\hbar \  dq}  \cr &+& {{\alpha_2} \over \sqrt{2
\pi \hbar (2E(n,p_c)
-2V(\tau))^{1/2}}}e^{-i\int_0^{\tau} \sqrt{2E(n,p_c) - 2V(\tau)}/\hbar \
dq}
\end{eqnarray}
for $\langle q_c|q'_c\rangle = \delta (q_c- q'_c)$.

Note that we have taken the normalization of $|p_c\rangle$ to
be $\epsilon$-dependent and that, in fact, \ref{cn}
diverges at $\epsilon = 0$.  This is because
$\epsilon$ is just the amount that the counter advances after
each swing of the pendulum and we will always read our clock
when the counter is in some window of width $\epsilon$ using the
operators \ref{obs}.  Thus, we will be
interested only in matrix elements which are proportional
to $\langle p'_c|\chi_a|p_c\rangle$. For
future reference, we note that this is
\be
\langle p'_c|\chi_a|p_c\rangle =  e^{i[a+1/2]\epsilon (p_c -
p'_c)/\hbar} \Bigl[ {{2 \hbar \sin [(p_c - p'_c)
\epsilon \hbar/2]} \over {\epsilon (p_c - p'_c)}} \Bigr].
\ee
which takes the value $1$ at $p'_c = p_c$ and decays
away from $p_c = p'_c$ with a width of $2/\hbar \epsilon$.

Furthermore, we will typically evaluate such matrix elements for
$p_c$ and $p'_c$ for which there are corresponding $n$ and $n'$
such that $E(n,p_c) = E(n',p'_c)$.  Thus, if the energy
levels of $H_p$ are separated by a spacing $\Delta E$, we have
$(p_c - p'_c)\epsilon \sim \Delta E / C^2$ from \ref{slope}.  But
semiclassically we expect $C \sim \sqrt{\hbar}/L$, where $L$ is the
distance between the turning points of $V$.  Thus, the function in
brackets is typically evaluated for $\Delta p \epsilon /\hbar
\sim \Delta E L^2/\hbar^2$.  For a smooth potential with $V(q)$
unbounded from above, this expression becomes large as we move to
high energy levels (see appendix \ref{delta}).
Thus, for this case we have $\epsilon \Delta p_c \hbar \gg 1$.
For our purposes then,
\be
\label{cme}
\langle p'_c|\chi_a|p_c\rangle =
\delta_{p_c,p'_c},
\ee
where \ref{cme} contains the Kronecker delta function.

\section{The Properties of Time}
\label{derive}

In this section, we derive the external time formalism from the
almost local formalism in the ideal clock limit.  We address the
external time formalism in three stages, first showing that the operators
follow a unitary law of evolution in \ref{unit}, then showing in
\ref{alg} that the operators
have the usual algebraic structure, and finally deducing in
\ref{map} the
explicit map between ${\cal H}_{phys}$ and the Hilbert space ${\cal
H}_{ext}$ of external
time quantum mechanics for the non-clock degrees of freedom.
Note that, by construction, ${\cal H}_1$ is isomorphic to ${\cal
H}_{ext}$.  Along the way, we will show that `conservation
of existence,' and thus conservation of probability in the external
time framework follows from unitarity and the algebraic structure.

\subsection{Unitarity}
\label{unit}

The first topic we address in our derivation of the external time
formalism is the issue of {\it unitarity}.  We will show below that, in
the context of our approximations, we have
\be
\label{ev}
A_{\tau + \delta} = e^{i\delta P_\tau} A_\tau e^{-i \delta P_\tau}
\ee
when $A = A(Q^i_\perp, P_j^\perp)$ is built from the non-clock
degrees of freedom.  Since, as
was noted in section \ref{AIC},
$P_\tau$ defines a symmetric operator
on ${\cal H}_{phys}$,
it follows that \ref{ev} describes unitary evolution in the clock time
$\tau$.

In a small digression, we note that there is an interesting case for
which the evolution in clock time is {\it exactly} unitary but
is not exactly of the form \ref{ev}.  This is the case in
which the clock momentum $P$ is conserved ($[P,H] = 0$) so that
$P(t) = P$ is independent of $t$ and defines a physical operator.
Using this property, a brief study of \ref{obs} shows that $A_{\tau + \delta}
= e^{i \delta P} A_\tau e^{-i \delta P}$ and we have exact unitary
evolution.  However, $P$ and $P_\tau = P\openone_\tau$ are different
operators unless we have $\openone_{\tau} =
\openone$ which, as appendix \ref{fleet} describes, is not the case
for interesting such systems.  Note also that
$i\hbar { \p \over {\p \tau}}
\openone_\tau = [\openone_\tau,P]$ may not vanish.
This highlights the difference between
unitarity of evolution and the property ${\p \over {\p \tau}}
\openone_\tau = 0$ which we shall call `conservation of existence'
and which is related to the more familiar conservation
of probability in external time quantum mechanics (see \ref{map}).
Unitarity
of evolution is associated with invariance of the {\it system} under
shifting the zero of clock time.  On the other hand, conservation of
existence in some state is associated with the fact that, in that
state, the pendulum moves through
not only the value $\tau$, but
also all nearby values of $q$.  Due to geodesic incompleteness,
this will fail to hold, for
example, when a clock that measures metric-defined
proper time finds itself near a spacetime singularity.
However, {\it both}
unitarity and conservation of existence will follow in our `ideal clock
limit'.

We now proceed to derive \ref{ev} using the scheme introduced in
\ref{approx}.
Our strategy will be to compare matrix elements of $[A_\tau,P_\tau]$
with those of $-i \hbar {\p \over {\p t}} A_\tau$ and to show that they
agree to within the desired accuracy.

The first step in this process will be to evaluate matrix elements
of $P_{\tau}$ in a convenient basis.
To this end, consider the states $|n,p_c, E_1, a \rangle
= |n,p_c\rangle \otimes |E_1,a\rangle$ where
$\langle E_1,a |E'_1, a'\rangle = \delta_{a,a'} \delta^{(?)}(E_1,E'_1)$
and $\delta^{(?)}$ is either a Dirac or a
Kronecker delta-function as is appropriate to the spectrum of $H_1$.
As before, the physical `subspace' is constructed from the
states for which $E_1 = - E(n,p_c)$.

For later convenience, we consider not only $P_\tau$, but also
the operators $(P^m)_\tau$ for $m \in \{0\} \cup {\bf Z}^+$
defined by \ref{obs} with $A = P^m$.
Thus, using \ref{mes} and
the above basis of states we have
\begin{eqnarray}
\label{pm}
(P^m)_{\tau} &|&n, p_c(n,-E), E, a \rangle
= \pi i \sum_{n',E',a'}  |n', p_c(n',-E'), E', a' \rangle \cr
&\times& \langle n', p_c(n',-E'), E', a'|\theta(P) \chi_0
\{ P^m, [P^2/2, {\p \over {\p t}}\theta(Q-\tau)]\} \theta(P)
|n, p_c(n,-E), E, a \rangle
\end{eqnarray}
up to terms of order $\epsilon p_c$ where $p_c(n,-E)$ is the value
of $p_c$ for which $-E = E(n,p_c)$.  Note that our choice of
normalization \ref{norm} was crucial in deriving \ref{pm}.
But, using \ref{cme}\footnote{Recall that \ref{cme} is valid only
when $V$ is unbounded from above.  This assumption is not actually
needed to derive that \ref{ev} holds in the form
$\langle \phi| -i\hbar {\p \over {\p \tau}} A_{\tau} | \psi \rangle
= \langle \phi| [P_\tau, A_\tau]| \psi \rangle$
for states with $\epsilon p_c \ll 1$ as the
cross terms removed by using \ref{cme} will be irrelevant in this case.
However, in the interests of a simpler presentation, we use \ref{cme}
here.} and the
fact that the clock coordinates and momenta commute with the
canonical variables of the other degrees of freedom, this
becomes
\begin{eqnarray}
\label{pt1}
P^m_{\tau} |n, p_c(n,-E), E, a \rangle
&=&  \pi i |n, p_c(n,-E), E, a \rangle \cr &\times&
  \langle n; p_c(n,-E) |\theta(P)
\{ P^m, [P^2/2, \theta(Q-\tau)]\} \theta(P)
|n; p_c(n,-E) \rangle
\end{eqnarray}
Let $\Pi_\tau = |\tau\rangle \langle \tau |$ so that the commutator in
\ref{pt1} becomes $[P^2/2,\theta(Q-\tau)] =  -{i \over 2}  \{P,
\Pi_\tau\} \hbar$.  Thus, we need to evaluate the matrix
elements $\langle \tau | P^k \theta(P) |n;p_c\rangle$ for
$0 \leq k \leq m$.

To do so, we again consider \ref{WKB} and note that, to the desired
accuracy, the projection $\theta(P)$ acting on $|n;p_c\rangle$
yields just the first term ($\langle \tau|n;p_c\rangle_+$)
in \ref{WKB} and the projection $\theta(-P)$
yields the second ($\langle \tau|n,p_c\rangle_-$).
This follows since the negative (and zero) frequency
components of $|n;p_c\rangle_+$
are given by Fourier transform of the wavefunction
$\langle \tau | n;p_c\rangle_+$ and the integral which defines
this transform may be divided into two parts.
The first part comes from the integral over the region between the
turning points and, as it is given by the Fourier transform of \ref{WKB}
in this region, vanishes faster than any power of $E$ by
the Riemann-Lebesgue theorem.  The other contribution comes from
the region within a distance of order $\lambda_0 = \hbar/\sqrt{E}$ of the
turning points (the contribution from outside
the turning points is negligible).  In this region, the magnitude of the
wavefunction may be estimated from the value of \ref{WKB} a distance
$\lambda_0$ from the turning point and is roughly $\sqrt{\lambda_0}/\hbar$.
Thus, this term is of order $\lambda_0^{3/2}/\hbar$.  In contrast, the
largest positive frequency component is of order
$\sqrt{\lambda_0}L/\hbar$, where $L$ is the distance between the turning
points.  Thus, the negative and zero frequency parts of $|n;p_c\rangle_+$
are of order $\lambda_0/L \sim 1/n$ which is negligible when $n$ is large
enough that we may take $\tau$ far from a turning point.

Now, to lowest order in $\hbar$, we compute
$\langle \tau| P^k \theta(P) |n;p_c\rangle
 = (2E(n,p_c) - 2V(\tau))^{k/2} \langle \tau| n;p_c\rangle_+$ from
\ref{WKB}.
Since $|\langle \tau|n;p_c\rangle_+|^2 = 1/(2\pi \hbar [2E(n,p_c) -
2V(\tau)]^{1/2})$, it follows that
\be
\label{pt}
P^m_\tau |n,p_c, -E(n,p_c),a \rangle = ( 2E(n,p_c) - 2V(\tau))^{m/2}
|n, p_c,-E(n,p_c),a \rangle
\ee
to the accuracy we consider.
Note that this holds even when $m=0$.  Thus, the
existence operator $\openone_{\tau}$ is the identity to this
approximation and, in particular, ${\p \over {\p \tau}}
\openone_{\tau} = 0$ up to terms of order $\hbar$ so that we have
derived `conservation of existence.'

Such results are no surprise; they are
exactly what should be expected from a semiclassical approximation to
a system of the form \ref{form}. In fact, there
have been attempts to quantize reparametrization invariant
systems {\it beginning} with an equation of this form \cite{GD,BK,SL}.
That this equation can be
derived from the basic formalism of \cite{QORD} is another
verification that this approach is physically reasonable.
Also, working within this formalism, we see that the issue of
`what happens when the argument of the square root becomes negative'
is a non-physical issue.  In this regime, the approximations considered
are not valid and, from general arguments, we have no reason to expect
that our pendulum should function well as a clock.

We may now use the result \ref{pt} to compute matrix elements of
$[P_\tau,A_\tau]$:
\begin{eqnarray}
\label{CAP}
\langle n',&p'c,&-E(n',p'_c), a'|[P_\tau, A_\tau] |n,p_c,
-E(n,p_c), a\rangle_{phys} = (\sqrt{2E(n',p'_c) - 2V(\tau)} -
\sqrt{2E(n,p_c) - 2V(\tau)}) \cr &\times&
\langle n',p'c, -E(n',p'_c), a'|A_\tau
|n,p_c, -E(n,p_c), a\rangle_{phys}
\end{eqnarray}
to lowest order in $\hbar$.  The subscripts $phys$ serve as reminders of
the Hilbert space in which the inner products are to be taken.
  We note that there is some subtlety
in concluding the above result as the {\it absolute} accuracy of
\ref{CAP} is higher than that of the expression
for $\langle \phi | A_\tau P_\tau |\psi\rangle$ or $\langle \phi |
P_\tau A_\tau|\psi\rangle$.
The point is that, because this expression is a
commutator, we expect that \ref{CAP} is ${\cal O}(\hbar)$.
Nevertheless, we may conclude \ref{CAP} to the desired order as the
corrections to \ref{pt} will be of the form $\hbar f(P,Q)$, whose
commutator with $A_\tau$ will be subleading, of order $\hbar^2$.

For comparison, we now compute matrix elements of the time
derivative $-i\hbar {\p \over {\p \tau}} A_{\tau}$.  Applying the
same general formalism and making use of the tensor product nature
of our basis states we find
\begin{eqnarray}
-i \hbar \langle n',&p'_c&, -E(n',p'_c), a'| {\p \over {\p \tau}}
A_\tau | n, p_c, -E(n,p_c), a \rangle_{phys} \cr &=& i \pi \hbar^2
{\p \over {\p \tau}}\langle n';p'_c| \theta(P) \{P,\Pi_\tau\}
\theta(P) |n;p_c\rangle \langle -E(n',p'_c), a'|A|-E(n,p_c), a\rangle.
\end{eqnarray}
Using \ref{WKB}, this agrees with \ref{CAP} to leading order in
$\hbar$.  When $V(Q)=0$, this result and \ref{pt} are exact since there
are no corrections to the WKB approximation and the counter is
irrelevant
so that we may set $\epsilon p_c =0$.
Thus, within our approximation scheme we
have $[P_\tau, A_\tau] = - i \hbar {\p \over {\p \tau}} A_\tau$ on
${\cal H}_{phys}$, verifying \ref{ev}.  Since $P_{\tau}$
is symmetric on ${\cal H}_{aux}$, we have shown that evolution in
the parameter $\tau$ is unitary, modulo issues of convergence of the
integrals defining $P_\tau$ and $A_\tau$ and of the domain of
definition of $P_\tau$.  This is the desired result.

Note that while the same calculation \ref{pt} which showed the unitarity of
evolution allowed us to derive conservation of existence
(${\p \over {\p \tau}} \openone_\tau = 0$), this result does {\it not}
follow directly from unitarity.  This is because unitarity by
itself does not provide a way to compute the commutator
$[P_\tau, \openone_\tau]$.  We again recall (see appendix \ref{fleet})
that for a homogeneous comoving fleet of ideal metric
clocks in a homogeneous cosmology, unitarity may hold exactly but
we will not, in general, have conservation of existence.

\subsection{The Equal Time Algebra}
\label{alg}

We now turn to the second feature of external time quantum mechanics
and show that it, too, follows from our formalism.  Here, we consider
the algebraic structure of the operators $A_\tau$ and, in particular,
show that if $Q^i_\perp$ and $P_j^\perp$ are the canonical coordinates and
momenta of the non-clock degrees of freedom, then $(Q^i_\tau)_\perp$ and
$(P_j^{\perp})_\tau$ satisfy the canonical commutation relations to lowest
order in $\hbar$.  In general, we verify that $\langle
\phi|[A_\tau, B_\tau] | \psi \rangle
= \langle \phi| ([A,B])_\tau | \psi \rangle
(1 + {\cal O}(\hbar) + {\cal O}(\epsilon p_c))$
for $A = A(Q^i_\perp,(P_j)_\perp)$
and $B = B(Q^i_\perp,P_j^\perp)$ and any two states $|\phi \rangle$
and $|\psi \rangle$ that satisfy the requirements of our approximation
scheme.
{}From \ref{obs} we also have $(A + B)_\tau = A_\tau + B_\tau$, and we
will see that the argument given below is also sufficient to
derive $\langle \phi |(AB)_\tau |\psi \rangle
= \langle \phi| A_\tau B_\tau | \psi \rangle$ to lowest order in $\hbar$.
Thus, to this accuracy the map ${}_\tau: A \mapsto A_\tau$
preserves the entire equal time algebraic structure of external time
quantum mechanics.  Together with the results of \ref{unit},
this in fact implies preservation to this order
of the algebraic structure associated
with all Heisenberg operators and the Hamiltonian $\sqrt{-2H_1
-  2V(\tau)}$.

Below, we will once again see that the presence of the `counter'
is critical.  That this should be so can be seen even in the
calculation of the classical Poisson Brackets and, in fact, the
calculation below is essentially the same as the classical one.
This is because we work in the
semiclassical approximation only to lowest order in $\hbar$ -- the
same order to which the ordering of factors may be ignored.
Nonetheless, the calculation below is presented in quantum
form to show that all of the ${\cal O}(\hbar)$ corrections
come from ignoring the ordering of {\it clock} variables only.
Operators associated with the other degrees of freedom are treated
in a fully quantum manner so that we expect the terms ignored to
be negligible in the ideal clock limit.

It will be easiest to compute the commutator of $A_\tau$ with
$B_\tau$ in the nonphysical space ${\cal H}_{aux}$ and then take this
result over to the physical space.  Thus, we must calculate
the integrals
\begin{eqnarray}
\label{cint}
[A_\tau,B_\tau] = \int_{-\infty}^{\infty} dt \int_{-\infty}^{\infty}
dt' & \Big[\theta(P(t)) A(t) \chi_0(t) {\p \over {\p t}} \theta(Q(t) - \tau)
\theta(P(t)), \cr
& \theta(P(t')) A(t') \chi_0(t') {\p \over {\p t'}} \theta(Q(t')
- \tau) \theta(P(t'))\Big]
\end{eqnarray}
where $A(t) = A(Q^i_\perp (t),P_j^\perp (t))$ and $B(t) =
B(Q^i_\perp
(t),P_j^\perp(t))$ are independent of the clock variables.
We will not attempt to explicitly expand the commutator in
\ref{cint} but, instead, we simply note that the derivation law
$[AB,C] = A[B,C] + [A,C]B$ can be used to express this commutator
as a sum of four kinds of terms,
depending on whether the $\theta(P)$, $\chi_0$, and
$\theta(Q-\tau)$
factors appear inside or outside the commutator.  We then treat terms of
each type separately.

We first consider the term in which both $ \theta(P(t))
\chi_0(t) {\p \over
{\p t}} \theta(Q(t) - \tau)$ and  \break $ \theta (P(t'))\chi_0(t') {\p \over
{\p t'}} \theta(Q(t') - \tau)$ appear {\it outside} of the commutator.
For this term, we use the semiclassical approximation to
write ${\p \over {\p t}} \theta(Q(t) - \tau) =
N(t) P(t) \Pi_\tau(t) + {\cal O}(\hbar)$ and similarly for
$t'$.  Again ignoring the commutator of clock variables, move
$\Pi_\tau(t) \theta(P(t)) \chi_0(t) N(t') P(t')
\Pi_\tau(t') \chi_0(t')$ to the
right.  These operators will now act directly on a state $|\psi\rangle$
in which the
clock behaves semiclassically.
Thus, we consider an expression of the form
\be
\langle \tau,q_c,q_\perp| \theta(P(t)
\chi_0(t) N(t') P(t')  \Pi_\tau(t') \chi_0(t') \theta(P(t))
|\psi\rangle
\ee
where  $|\tau, q_c, q_\perp \rangle = |\tau \rangle
\otimes |q_c \rangle \otimes |q_\perp \rangle$ and $|q_\perp \rangle$
denotes an arbitrary basis of ${\cal H}_1$.
We now use the fact that a semiclassical expectation value of the
above form may
be evaluated by considering the values of the corresponding classical
quantities along a classical path for which $q(t)=\tau$.
Note that along {\it any} such  path for which $p(t), p(t') >0$ and
both $q_c(t)$ and $q_c(t')$ lie in the interval $[0,\epsilon]$,
we have\footnote{Strictly
speaking, this is true only when the lapse is chosen to be everywhere
positive.  That the actual result is equivalent to this for arbitrary
$n(t)$ follows from the fact that the operator $[A_\tau,B_\tau]$
is independent of the choice of $n(t)$, though it may be derived
directly as well.} $N(t') p(t') \delta (q(t') - \tau) = \delta(t-t')$.
Thus, we
may replace the corresponding set of operators by this
expression and perform the integral over $t'$.  The
remaining commutator is just $[A(t) \theta(P(t)), B(t)\theta(P(t))]
= \theta(P(t)) [A, B](t)$ so that this term yields just
the desired result: $([A,B])_\tau$.

Clearly, we must show that all other terms generated are negligible.
We will now address type II terms, where the commutator is of the
form $[ {\p \over {\p t}} \theta(Q(t) - \tau), {\cal A}(t')]$
for some ${\cal A}(t')$ and where $\chi_0(t')$ appears outside
the commutator.  Using our semiclassical approximation, we
may replace this commutator with
\be
{\p \over {\p t}} \Big( \delta(Q(t) - \tau) [Q(t), {\cal A}(t')] \Big)
\ee
and integrate by parts with respect to $t$.  The factor of
$\chi_0(t) \delta(Q(t) - \tau)$
guarantees that the boundary terms vanish.  We now reorganize
the terms as above, with, perhaps, $\chi_0(t)$ being replaced
by ${\p \over {\p t}} \chi_0(t) \cong
{\p \over {\p t}}Q_c(t) (\delta(Q_c(t)) - \delta(Q_c(t) -\epsilon))$
or $\theta(P(t))$ being replaced by ${\p \over {\p t}} \theta(P(t))
\cong {\p \over {\p t}}P(t)
\delta(P(t))$ in some terms due to the action of ${\p \over {\p t}}$
in the integration by parts.
Even with such replacements, for $n(t) > 0$ we have
${\p \over {\p t}}
\theta(q(t') -\tau) = \delta(t-t')$ on all relevant classical solutions
with $q(t) = \tau$.  Thus, we may again replace the
corresponding operator with $\delta(t-t')$ so that the commutator
is evaluated at equal times.  Note that the only factors
which may be present on the right side of the commutator which do not
commute with the $Q(t)$ on the left are the $\theta(P(t))$.  But
$[Q(t), \theta(P(t))] = i \hbar \delta(P(t))$ and
$\langle \tau| \delta(P) |\psi\rangle$ is as small as the negative frequency
components of our WKB states; that is, it is negligible by
the argument of section \ref{unit}.

Terms of the third type are just those terms obtained from type II
terms by interchanging $t$ and $t'$ (which may be handled in the
analogous way) and the term containing $[{\p \over {\p t}}
\theta(Q(t) - \tau), {\p \over {\p t}} \theta(Q(t') - \tau)]$.
This term is shown to be negligible by integrating
by parts with respect to both $t$ and $t'$ and using arguments much like
those above to set $t=t'$, whereupon the commutator vanishes.

To deal with the remaining terms, we note that
\be
[\chi_0(t), {\cal B}] = [Q_c(t), {\cal B}] (\delta (Q_c(t)
- \delta (Q_c(t) - \epsilon))(1 + {\cal O}(\hbar))
\ee
and
\be
[\theta(P(t)), {\cal B}] = [P(t), {\cal B}] (\delta (P(t))
(1 + {\cal O}(\hbar)).
\ee
The corresponding terms then vanish by the assumption
that $|\psi\rangle$ has negligible component for which both
$q = \tau$ and $q_c =0 $
or $q_c = \epsilon$ and by the argument of \ref{unit} that the zero
and negative frequency components of $|\psi\rangle_+$ are
negligible.

We thus find that, when acting on a state satisfying our
requirements, we have
$[A_\tau, B_\tau] = ([A,B])_\tau$ to leading order
in $\hbar$, as expected.  In particular, $(Q_\perp^i)_\tau$
and $(P^\perp_j)_\tau$ satisfy the canonical commutation relations.
We emphasize that the terms
ignored come from commutators of clock variables so that they may be
expected to be small whenever the clock, but not necessarily the
entire system, behaves semiclassically.  While it does not follow
from the above argument, the results of section IVB of
\cite{QORD} show that  when $V(Q)=0$ and $H_1 = \sum_{ij} g^{ij}
P^\perp_i P^\perp_h$ for any $g^{ij} \in {\bf R}$, the
$(Q^i_\perp)^+_\tau$ and the $(P^\perp_j)^+_\tau$ are just the usual
positive frequency Newton-Wigner operators and so satisfy the
canonical commutation relations {\it exactly}.  This is a consequence
of the particular operator ordering chosen in \ref{obs}, even though
the choice of ordering does not effect the above semiclassical
derivation.
Finally, we note that the arguments applied to the type I terms
above can also be used
to show that $A_\tau B_\tau  = (AB)_\tau$ to leading order in $\hbar$
so that, to this order, the full algebra associated with external
time quantum mechanics has been reproduced.

In passing, we note that had we not explicitly shown `conservation of
existence' in the previous subsection, the results of \ref{unit}
combined with the above {\it would} have been sufficient to derive this
feature.  That is,
conservation of existence follows from the presence of a
reasonable generator of time translations and from the usual
algebraic structure of the theory.  The derivation proceeds as
follows.  Note that if the operator $A(t)$ is
independent of clock variables and is a constant of motion,
then $A_\tau = A \openone_\tau$.  Thus, $[\sqrt{-2H_1 - 2V(\tau)}]_\tau
= (-2H_1 - 2V(\tau))^{1/2} \openone_\tau$.  But this is just
$P_\tau \openone_\tau = P_\tau$ by the above.  We may calculate
the commutator of the existence operator with the time translation
generator as $[P_\tau, \openone_\tau] = [(\sqrt{-2H_1 - 2V(\tau)})_\tau,
\openone_\tau]$ and, since $[\sqrt{-2H_1 - 2V(\tau)}, \openone] = 0$, this
expression vanishes by the result just derived.

\subsection{The Map on States and the Wheeler-DeWitt Equation}
\label{map}

We now complete our derivation of external time
quantum mechanics by constructing the explicit map between our physical
states and states in the space ${\cal H}_{ext} \cong {\cal H}_1$.
To begin, we consider the
expectation value of an operator $A_\tau$ in some
state $|\psi\rangle$ that satisfies the requirements of our
approximation scheme.

Rewriting the usual commutator $[H, \theta(Q(t) - \tau)]$ as
$-i \hbar \sqrt{P(t)}\Pi_\tau(t)\sqrt{P(t)}$ up to
${\cal O}(\hbar^2)$ corrections, we have

\be
\label{exp}
\langle \psi | A_\tau|\psi\rangle_{phys} = 2 \pi \hbar \langle \psi|
\theta(P) \sqrt{P} \chi_0 \Pi_\tau A \sqrt{P} \theta(P)|\psi\rangle_{aux}
\ee
where all of the operators on the right hand side are evaluated
at the parameter time $t = 0$ and the subscripts $phys$ and $aux$
serve as reminders of the appropriate Hilbert space in which each inner
product is to be taken.

Expressing our states as $|\psi\rangle =
\sum_{n,E,a} \psi(n,E,a)
|n, p_c(n,-E), E, a\rangle$,
equation \ref{exp} may be written
as $Tr[A \rho(\tau)]$ where $\rho(\tau)$ is a
density matrix associated
with the Hilbert space ${\cal H}_1$ and is defined by:
\begin{eqnarray}
\label{rho}
\langle E',a'|\rho(\tau)|E,a\rangle = \sum_{n,n'}
\psi^*(n',E',a') \psi(n,E,a) && \langle n';p_c(n',-E') | \theta(P)
\sqrt{P}| \tau \rangle \cr
&\times &  \langle \tau| \sqrt{P} \theta(P) | n;p_c(n,-E)
\rangle \langle p_c| \chi_0 | p'_c\rangle.
\end{eqnarray}
Note that the $\tau$ dependence of \ref{rho} is exactly that of
a Schr\"odinger picture density matrix that evolves in time
under the Hamiltonian $\sqrt{-2H_1 - 2V(\tau)}$ since
$-i \hbar {{\p \rho} \over {\p \tau }} = [\rho(\tau), \sqrt{-2H_1
- 2V(\tau)}]$.
Thus, \ref{rho} defines a map from states in ${\cal H}_{phys}$
to density matrices in the Schr\"odinger picture of external time
quantum mechanics.  In fact, \ref{rho} would describe a pure state
except for the factor $\langle p_c |\chi_0 | p'_c\rangle$.

Recall, however, that we consider the limit where the coupling
to the counter is small
($p_c \epsilon /\hbar
\ll 1$) so that the bands produced by coupling the counter
to the pendulum are well separated.  In particular, we
consider only
those states $|\psi\rangle$ for which the support of $\psi(n,E,a)$
is concentrated on values of $n$ and $E$ for which $|p_c(n,-E)| \leq
\beta$ for some $\beta \epsilon/\hbar  \ll 1$.  Since $\langle
p_c |\chi_0 |p'_c\rangle$ is a distribution of unit height and
width $2 \hbar/\epsilon$, this factor is essentially one in \ref{rho}
when $\psi(n,E,a)$ is nonzero and $\rho(\tau)$ effectively
describes a pure state.

For our states then, it is useful to consider the projections
\begin{eqnarray}
\label{eta}
&\eta:& {\cal H}_{aux} \rightarrow {\cal H}_p \otimes {\cal H}_1 \cr
&&|\psi\rangle \mapsto \sum_{n,E,a} \psi(n,E,a) |n,E,a \rangle
\end{eqnarray}
and
\begin{eqnarray}
\label{sigma}
&\sigma_{\tau}:& {\cal H}_p \otimes {\cal H}_1 \rightarrow {\cal H}_1
\cr && |\phi\rangle = \int d\tau' \sum_{q_\perp} \phi(\tau', q_\perp)
|\tau', q_\perp \rangle \mapsto \sum_{q_\perp} \phi(\tau, q_\perp) |q_\perp
\rangle
\end{eqnarray}
where $|\tau, q_\perp \rangle  = |\tau \rangle \otimes
| q_\perp \rangle$ and $|q_\perp \rangle$ is an arbitrary basis of ${\cal
H}_1$.
We may then write $\rho(\tau) = |\psi;\tau\rangle \langle \psi;\tau|$
where
\be
\label{pem}
|\psi;\tau\rangle = \sigma_{\tau} \theta(P) \sqrt{P} \eta |\psi\rangle
\ee
so that we have derived $\langle \psi| A_\tau | \psi \rangle =
\langle \psi; \tau|A| \psi; \tau \rangle$.  Similarly,
for two such states $|\phi\rangle$ and $|\psi\rangle$, we have
\be
\label{exp2}
\langle \phi|A_\tau|\psi\rangle = \langle \phi;\tau |A|\psi;\tau \rangle.
\ee
The state $|\psi;\tau\rangle$ satisfies
\be
\label{Schro}
- i \hbar {\p \over {\p \tau}} |\psi ; \tau \rangle =
\sqrt{-2H_1 - 2V(\tau)}|\psi;\tau
\rangle
\ee
so that we may interpret $|\psi;\tau \rangle$ as a Schr\"odinger
picture state in ${\cal H}_{ext} \cong {\cal H}_1$
with Hamiltonian $\sqrt{-2H_1 - 2V(\tau)}$.  Note that we arrived at
a Schr\"dinger picture description due to the implicit mapping of the
operator $A_\tau$ to the operator $A(t=0)$ in \ref{exp} and
\ref{exp2}.  Had
we used the Heisenberg picture operator $A_H(t = \tau)$ associated with
$A$ by the Hamiltonian $\sqrt{-2H_1 - 2V(\tau)}$, we would have arrived
at a Heisenberg picture description.

Note that the map \ref{pem} takes a familiar form.
In particular,
for $A = \openone$, \ref{exp2} is just
\begin{eqnarray}
\label{eip}
\langle \phi | \openone_\tau | \psi \rangle_{phys}
&=& ( \sigma_{\tau} \theta(P) \eta(P) |\phi \rangle,
\sqrt{-2H_1 - 2V(\tau)} \sigma_\tau \theta(P) \eta |\psi \rangle ) \cr
&=& {{i \hbar } \over 2} ( {\p \over {\p \tau}} \sigma_\tau \theta(P)
\eta |\phi \rangle, \sigma_\tau \theta(P) \eta |\psi \rangle )_{{\cal
H}_1}
- {{i \hbar} \over 2}( \sigma_\tau \theta(P)
\eta |\phi \rangle, {\p \over {\p \tau}}
\sigma_\tau \theta(P) \eta |\psi \rangle )_{{\cal H}_1}
\end{eqnarray}
and, since $\openone_\tau = \openone$ in our approximation, we see
that up to constant factors the inner product on ${\cal H}_{phys}$
in the ideal clock limit is
the positive frequency part of the familiar Klein-Gordon
inner product\footnote{The author would
like to thank Jim Hartle for bringing this result to his attention
in the case where the potential $V(\tau)$ has no critical
points and is asymptotically constant.}.
Note, however, that the vector field ${\p \over {\p
\tau}}$ may be {\it either} spacelike {\it or} timelike with respect to
the metric defined by the kinetic terms in the constraint.
Similarly, the operators $A_\tau$ become the positive frequency
Newton-Wigner operators in the ideal clock limit and, in this
limit, our approach will agree with that of Wald \cite{HW}.
In particular, in contrast to the suggestions of \cite{DP,SCPI}
it follows that in this scheme the square of the
`wavefunction' ($\langle q_\perp, \tau | \eta |\psi \rangle |^2$)
may {\it not} be interpreted as a probability density for `the value of
$q_\perp$ when the clock reads $\tau$.'
Instead, this interpretation belongs to the quantity
$\langle q_\perp | \sigma_\tau \sqrt{P} \theta(P) \eta |\psi \rangle |^2$.

A few comments on exact results are now in order.  From section IVB
of \cite{QORD}, it follows that any corrections to \ref{eip} vanish
when $V(Q) = 0$.  This is surprising from our derivation since we have
completely ignored the effects of the commutator $[P,\Pi_\tau]$.
However, when $V(Q) = \tau$ the WKB approximation is exact and
$P \theta(P) |\psi\rangle = \sqrt{-H_1} \theta(P) |\psi\rangle$.
Thus, when $[A,H_1] = 0$, we have $\langle \phi| \theta(P)
A [P,\Pi_\tau] \theta(P) |\psi\rangle = 0$ so that \ref{exp2}
is exact.  In particular, this follows for $A = \openone$.  When
$H_1 = \sum g_{ij} P^\perp_i P^\perp_j$, direct calculation as in
section IVB of \cite{QORD} show that \ref{exp2} holds exactly for
$A = Q^i_\perp$ as well.

Finally, note that \ref{eip} may also be written as
\be
\langle \phi| \openone_\tau |\psi \rangle_{phys} = \langle \phi; \tau|
\psi ; \tau \rangle_{{\cal H}_1}.
\ee
Differentiating both sides with respect to $\tau$, we see
that conservation of probability (that is, of the right hand side)
is an explicit consequence of conservation of existence on the left.

\section{Discussion}
\label{diss}
We have now completed our derivation of the three defining properties of
external time quantum mechanics.  Recall that the entire formalism
followed using only the approximations of \ref{approx} that we have an
ideal limit of an almost ideal clock as described in \ref{AIC}.
In particular, in contrast to
standard derivations\cite{Banks,Jim,JH,CK1,CK2}
of evolution through the use of the WKB
approximation, our assumptions do {\it not} imply that the clock is
large or massive so that the non-clock degrees of freedom may, in this
approach, represent more than a small perturbation and the clock need not
be in a state with nearly zero energy.
In addition, the complete formalism
of \cite{QORD} has allowed a more extensive treatment than was
previously possible -- we were able to address the issue of unitary
evolution in the {\it physical} inner product and to derive the
complete map from the physical Hilbert space to the corresponding
Hilbert space of external time quantum mechanics\footnote{Note that, despite
the introduction of an inner product in, for example, \cite{Jim,JH}
for the external time system, the complete map from, in this case,
solutions of the Wheeler-DeWitt equation to ${\cal H}_{ext}$ was not
provided.  Instead, this method can only define the required
map for a collection of
states with a  common semiclassical prefactor due to the
possible addition of a constant to $S$ of \cite{Jim} and the ambiguity
in the normalization of the factor $C$ of \cite{JH}
which are natural
consequences of the lack of a physical inner product.  However, within
such a class of states, \ref{eip} agrees with \cite{Jim,JH} up to an
overall normalization.}.

That such a derivation was possible may be considered a surprise in
light of the results of \cite{WS,HC} which demonstrate bounds on the
extent to which `realistic clocks' (including the free particle) may
approximate the ideal clock of \ref{ideal}.  Note however that the
point made in \cite{WS,HC} is that dispersion effects limit the
extent to which the readings of physical clocks may correspond to proper time.
Thus, from
the above results and the facts that our approximations are exact for a
system of free particles and that \ref{unit} and \ref{eip} are exact
whenever the clock is a free particle,
we may conclude that the use of a clock which measures
proper time in a state-independent way
is not fundamental to a derivation of unitary evolution,
the canonical commutation relations, or the well-known Hilbert space
structure.

However, despite arriving at the expected formalism, the final
result was {\it not} quantum mechanics in the usual form.  Indeed,
where we would expect to find $-H_1$ as the Hamiltonian in \ref{Schro},
we instead find $\sqrt{-2H_1 - 2V(\tau)}$.  This is a result of the
above mentioned dispersion.
Note that for $A = A(Q^i_\perp,
P^\perp_j)$, we have $[-H_1,A] = {1 \over 2} \{\sqrt{-2 H_1 -2 V(\tau)},
[\sqrt{-2 H_1 -2  V(\tau)},A]\}$ so that, if the state $|\psi\rangle$
has a reasonably well defined value $p$ of the pendulum momentum $P$,
this is just $\langle \psi | [-H_1,A] | \psi \rangle =  p
\langle \psi | [ \sqrt{-2H_1 - 2V(\tau)},A] |\psi \rangle$.  As
described in \cite{WS,HC}, this is just the case in which our
clock may be said to measure ($p$ times
the) proper time.
In practice, real clocks are only used to the
extent that they measure proper time and thus agree with each other;
that is, to the extent that the dispersion above may be neglected.
Since every test of time evolution makes use of some sort of dispersive
clock, we see that our results are in fact much stronger than what is
required to derive
\begin{equation}
-i\hbar {\p \over {\p T}} |\psi;pT\rangle  = -H_1|\psi;pT\rangle,
\ee
where $T = \tau/p$ is the proper time elapsed since the clock read
zero, to the extent that this is supported
by observations.

It is worthwhile to point out three further interesting features
of our derivation.  First, note that our discussion required no concept
of decoherence as in \cite{CK1,CK2} between states (or parts of states)
with different semiclassical prefactors.  This was the case
because the observables themselves impose the condition that the pendulum is
moving to the right through the position $\tau$.  Second,  the
signs of the kinetic terms in $H_1$ relative to $H_0$ have not been
specified, so that our clock may represent a mechanical clock as in
\cite{Banks} or the scale factor of a cosmological model as in \cite{JH}.
Third, despite the presence of $\sqrt{-2H_1 - 2V(\tau)}$ in \ref{Schro}
and the fact that $H_1$ may not be negative definite, we have no need
to concern ourselves as in \cite{BK,SL}
with the negative eigenvalues of $-2H_1 - 2V(\tau)$.
This is because \ref{pt} guarantees that any state with non-negligible
components along the corresponding eigenvectors will not be in the image
of the map \ref{pem}.

Finally, we would like to return to the discussion of \ref{AIC}
and to comment on the physical situations in which our ideal clock
approximation should be expected to hold.  Recall that our approximation
scheme was designed to guarantee that our clock acts like
an ideal one.  However, this ideal clock records the passing of
lapse-defined proper time, not metric-defined proper time.
Had
the clock instead measured metric-defined proper time (assuming that it is
coupled to the description of the gravitational field used in \cite{QORD}
or \cite{BIX}), our system would be described (see appendix \ref{fleet})
by a constraint of the form
\be
\label{mbc}
0 = e^{3\alpha} H_0 + H_1
\ee
instead of by \ref{form}.  Here, $\alpha = {1 \over 6}
\ln \det g$ and $g$ is the 3-metric on a homogeneous slice of a
homogeneous cosmological model.  However, the metric-based and
lapse-based notions of almost
ideal clock will agree when $e^{3 \alpha}$ may be treated as an adiabatic
factor.  We estimate when this is the case by comparing the
contributions coming from $e^{3\alpha}$ and $H_0$ in
a typical term of the parameter time derivative
${\p \over {\p t}}$ of $e^{3 \alpha}H_0$.  Consider, for example,
${\p \over {\p t}} \Big( {{e^{3\alpha} P^2 } \over 2} \Big)
= P^2 e^{3 \alpha} \Big( { { 3 \dot{\alpha}} \over 2} + { {\dot{P}}
\over P} \Big)$ where a dot denotes ${\p \over {\p t}}$.
Thus, the adiabatic approximation should hold when ${{\dot{P}} \over
{\dot{\alpha}}} \gg P$.  Using $ \hbar \dot{A} \propto [A,NH]$ and the
fact that $H$ is typically of a form (see \cite {QORD,BIX}) such
that $[\alpha,H] = \hbar P_{\alpha}$, this condition is
\be
{{V'(Q)} \over {P_\alpha}} e^{3 \alpha} \gg P
\ee
which we may expect to hold when $e^{3 \alpha}$ is large.  Thus,
these two notions of almost ideal clock agree for large spatial volumes.
Inserting the appropriate dimensional factors, this condition says
that the 3-volume $V$ of a homogeneous slice should satisfy
$V \gg \ell_p^3 {{m_p} \over M}$ where $m_p$ and $\ell_p$ are
the Plank mass and length respectively while $M$ is the mass of the
clock (that is, we replace $H_p$ by $P^2/2M + V(Q)$).

This is in fact the best that we can expect.  Note that, on classical
solutions in these models, when the spatial volume is small the clock is
near the initial or final singularity.  Thus, even if the clock
accurately measures metric-defined proper time all the way up to the
singularity, the the position of the
pendulum will change very little
before the end (or beginning) of the universe is reached and
conservation of existence should fail for a metric-based clock.
In this way it follows that
we should not expect the usual external time
formalism to hold for a metric-based clock near a spacetime
singularity.

\acknowledgments
The author would like to thank Steve Giddings,
Jim Hartle, Gary Horowitz, David Lowe, David Marolf, and
Mark Srednicki for helpful questions, discussions, and comments.
Thanks also to
Claus Kiefer for  useful conversations over several years.
This work was supported by NSF grant PHY90-08502.

\appendix

\section{Level Spacing for Unbounded Potentials}
\label{delta}

In this appendix we show that if $V$ is unbounded from above and
of the type considered in section \ref{approx} then
$\Delta E L^2 /\hbar^2$ diverges for large $n$, where $\Delta E$
is the spacing between the $n$-th and $(n+1)$-th energy level of
the Hamiltonian $P^2/2+V(Q)$ and $L$ is the distance between the
classical turning points associated with energy $E$.  As in the main
text, we use the approximation scheme of \ref{AIC}
to invoke semiclassical methods.
We proceed by assuming that $D = \Delta E L^2/\hbar^2$ is bounded
and showing that $V$ must in turn be bounded from above.

First, recall that in the semiclassical limit the energy levels
are determined by the Bohr-Sommerfeld quantization condition (see, for
example, \cite{LL}) which is
roughly:
\be
L_n \sqrt{2E_n} = n \hbar.
\ee
Equivalently, $E_nL_n^2/\hbar^2 = n^2$.  Thus,
\be
\label{inc}
\sqrt{2}{{L^2 \Delta E } \over {\hbar^2}} +  {{2 \sqrt{2} E
L \Delta L} \over {\hbar^2}}
\cong 2n \rightarrow \infty.
\ee
If $D$ is bounded by some constant $D_{max}$, \ref{inc} implies that
$EL \Delta L \sim  n \hbar^2/\sqrt{2} \sim 2 L \sqrt{E} \hbar$ and that
$\Delta L \sim \hbar / \sqrt{E}$.  Since, however,
$\Delta E \leq D_{max} \hbar^2/L^2$, we have
\be
V' \sim \Delta E/ \Delta L \leq D_{max} \hbar \sqrt{E}/2 L^2
\ee
so that for large $L$, $V \leq - const_1/L + const_2$.  But, from
\ref{approx}, $V$ must increase with $L$.  We conclude that $V$ is
bounded above.

\section{A fleet of Clocks}
\label{fleet}

While nowhere have we considered full gravity in this work, the
almost local
formalism is directly applicable \cite{QORD,BIX}
to the homogeneous cosmological
models of \cite{bj,atu,RySh}.  Thus, we now investigate the notion
of metric-based clocks in such a model.  In particular,
we show in this appendix that a homogeneous
comoving fleet of noninteracting such clocks in a homogeneous spacetime
is described by a constraint of the form $e^{3 \alpha} P + H_1$
and not of the form \ref{ideal}.

Let us consider a 3+1 spacetime containing a fleet of metric-based
clocks which admits a foliation by a family of spacelike
surfaces $\Sigma_t$ that are homogeneous in the sense of \cite{bj}
with regard to the metric variables and such that the clocks are evenly
distributed with respect to the metric on each slice.  Let us also
assume that the clocks are {\it comoving} in the sense that the
4-velocity field $U^{\alpha}$ of the clock fleet is proportional to the
normal field $n^{\alpha}$ of the family of slices $\Sigma_t$.

Now such a fleet of clocks is described (see \cite{BM}) by the formalism
for presureless dust presented in \cite{BK}.  Thus, from equations
3.14 and 3.19 of \cite{BK}, the contribution of this system to the
constraints associated with the ADM lapse and shift are
\begin{eqnarray}
H^D_a &=& PU_a \cr
H^D_\perp &=& \sqrt{P^2 + g^{ab}H_a^D H_b^D}
\end{eqnarray}
where $a$ is an abstract index on the slice $\Sigma$, $g^{ab}$ is the
metric on this slice, and $P$ is the momentum conjugate to the
clock reading $T$.  But, since $U^{\alpha} \propto n^{\alpha}$, we
have $U_a = 0 = H^D_a$ and $H^D_\perp = P$.  Thus, when coupled to
gravity the homogeneous system may be described by a single
Hamiltonian constraint $H_{ADM} = H^D_\perp + H^G_{ADM}$ associated
with the ADM lapse.  The form of $H_{ADM}^G$ may be found in
\cite{bj,atu,RySh} and, in the absence of the clock, is not of a form
that is appropriate to the almost local formalism (see
\cite{QORD,BIX}).  Even when the clock
is present, such a constraint is inconvenient for the almost local
approach because it does not naturally define a unique {\it
self-adjoint}
operator (see \cite{GD}).  As discussed in \cite{GD}, this is not merely
a technical difficulty but is a reflection of the fact that
the corresponding classical hamiltonian vector field is not
complete; that is, that this constraint
describes a system which forms singularities in finite proper time {\it
as defined by the lapse associated with the constraint.}  The
problem may thus be eliminated by rescaling the lapse and constraint as in
\cite{QORD,BIX} to use instead
\be
\label{B2}
0 = H = e^{3 \alpha} (P+ H^G_{ADM})
\ee
which does naturally define a self-adjoint operator on $L^2({\cal Q})$.
The associated lapse is then
$N = e^{-3 \alpha} N_{ADM}$.

On a typical integral curve of the classical hamiltonian vector field
corresponding to \ref{B2}, our clock will take on only a limited
range of values.  Thus, we do not classically have
conservation of existence (${\p \over {\p \tau}} \openone_\tau = 0$)
or $\openone_\tau = \openone$.
In the quantum theory, we expect these to hold
only to the extent that
$e^{3 \alpha}$ may be treated as an adiabatic factor;
nevertheless, the
results of \ref{unit} show that evolution in the clock time is
always exactly unitary.

\end{document}